\documentstyle [12pt,a4]{article}
\newcommand{\be}{\beta}
\newcommand{\de}{\delta}

\newcommand{\vep}{\varepsilon}

\newcommand{\la}{\lambda}
\newcommand{\La}{\Lambda}
\newcommand{\Lao}{\Lambda_0}

\newcommand{\vp}{\varphi}
\newcommand{\uvp}{\underline{\varphi}}
\newcommand{\uhvp}{\underline{\hat{\varphi}}}
\newcommand{\up}{\underline{p}}
\newcommand{\uk}{\underline{k}}
\newcommand{\vecp}{\vec{{p}}}

\newcommand{\Cll}{C^{\Lambda,\Lambda_0}}
\newcommand{\Lll}{L^{\Lambda,\Lambda_0}}
\newcommand{\Llol}{L^{\Lambda_0,\Lambda_0}}

\newcommand{\cLll}{{\cal L}^{\Lambda,\Lambda_0}}
\newcommand{\Dll}{{\cal D}^{\Lambda,\Lambda_0}}
\newcommand{\cLlln}{{\cal L}^{\Lambda,\Lambda_0}_{l,n}}
\newcommand{\Dlln}{{\cal D}^{\Lambda,\Lambda_0}_{l,n}}
\newcommand{\cLol}{{\cal L}^{0,\Lambda_0}}
\newcommand{\cLlol}{{\cal L}^{\Lambda_0,\Lambda_0}}
\newcommand{\Dlol}{{\cal D}^{\Lambda_0,\Lambda_0}}
\newcommand{\pa}{\partial}
\newcommand{\ti}[1]{\tilde{#1}}
\newcommand{\qed}{\hfill \rule {1ex}{1ex}\\ }
\newcommand{\eq}{\begin{equation}}
\newcommand{\eqe}{\end{equation}}
\newcounter{saveeqn}

\begin{document}
\message{reelletc.tex (Version 1.0): Befehle zur Darstellung |R  |N, Aufruf=
z.B. \string\bbbr}
%
%
\message{reelletc.tex (Version 1.0): Befehle zur Darstellung |R  |N, Aufruf=
z.B. \string\bbbr}
\font \smallescriptscriptfont = cmr5
\font \smallescriptfont       = cmr5 at 7pt
\font \smalletextfont         = cmr5 at 10pt
\font \tensans                = cmss10
\font \fivesans               = cmss10 at 5pt
\font \sixsans                = cmss10 at 6pt
\font \sevensans              = cmss10 at 7pt
\font \ninesans               = cmss10 at 9pt
\newfam\sansfam
\textfont\sansfam=\tensans\scriptfont\sansfam=\sevensans
\scriptscriptfont\sansfam=\fivesans
\def\sans{\fam\sansfam\tensans}
\def\bbbr{{\rm I\!R}} 
\def\bbbn{{\rm I\!N}} 
\def\bbbE{{\rm I\!E}} 
\def\bbbm{{\rm I\!M}}
\def\bbbh{{\rm I\!H}}
\def\bbbk{{\rm I\!K}}
\def\bbbd{{\rm I\!D}}
\def\bbbp{{\rm I\!P}}
\def\bbbone{{\mathchoice {\rm 1\mskip-4mu l} {\rm 1\mskip-4mu l}
{\rm 1\mskip-4.5mu l} {\rm 1\mskip-5mu l}}}
\def\bbbc{{\mathchoice {\setbox0=\hbox{$\displaystyle\rm C$}\hbox{\hbox
to0pt{\kern0.4\wd0\vrule height0.9\ht0\hss}\box0}}
{\setbox0=\hbox{$\textstyle\rm C$}\hbox{\hbox
to0pt{\kern0.4\wd0\vrule height0.9\ht0\hss}\box0}}
{\setbox0=\hbox{$\scriptstyle\rm C$}\hbox{\hbox
to0pt{\kern0.4\wd0\vrule height0.9\ht0\hss}\box0}}
{\setbox0=\hbox{$\scriptscriptstyle\rm C$}\hbox{\hbox
to0pt{\kern0.4\wd0\vrule height0.9\ht0\hss}\box0}}}}

\def\bbbe{{\mathchoice {\setbox0=\hbox{\smalletextfont e}\hbox{\raise
0.1\ht0\hbox to0pt{\kern0.4\wd0\vrule width0.3pt height0.7\ht0\hss}\box0}}
{\setbox0=\hbox{\smalletextfont e}\hbox{\raise
0.1\ht0\hbox to0pt{\kern0.4\wd0\vrule width0.3pt height0.7\ht0\hss}\box0}}
{\setbox0=\hbox{\smallescriptfont e}\hbox{\raise
0.1\ht0\hbox to0pt{\kern0.5\wd0\vrule width0.2pt height0.7\ht0\hss}\box0}}
{\setbox0=\hbox{\smallescriptscriptfont e}\hbox{\raise
0.1\ht0\hbox to0pt{\kern0.4\wd0\vrule width0.2pt height0.7\ht0\hss}\box0}}}}

\def\bbbq{{\mathchoice {\setbox0=\hbox{$\displaystyle\rm Q$}\hbox{\raise
0.15\ht0\hbox to0pt{\kern0.4\wd0\vrule height0.8\ht0\hss}\box0}}
{\setbox0=\hbox{$\textstyle\rm Q$}\hbox{\raise
0.15\ht0\hbox to0pt{\kern0.4\wd0\vrule height0.8\ht0\hss}\box0}}
{\setbox0=\hbox{$\scriptstyle\rm Q$}\hbox{\raise
0.15\ht0\hbox to0pt{\kern0.4\wd0\vrule height0.7\ht0\hss}\box0}}
{\setbox0=\hbox{$\scriptscriptstyle\rm Q$}\hbox{\raise
0.15\ht0\hbox to0pt{\kern0.4\wd0\vrule height0.7\ht0\hss}\box0}}}}

\def\bbbt{{\mathchoice {\setbox0=\hbox{$\displaystyle\rm
T$}\hbox{\hbox to0pt{\kern0.3\wd0\vrule height0.9\ht0\hss}\box0}}
{\setbox0=\hbox{$\textstyle\rm T$}\hbox{\hbox
to0pt{\kern0.3\wd0\vrule height0.9\ht0\hss}\box0}}
{\setbox0=\hbox{$\scriptstyle\rm T$}\hbox{\hbox
to0pt{\kern0.3\wd0\vrule height0.9\ht0\hss}\box0}}
{\setbox0=\hbox{$\scriptscriptstyle\rm T$}\hbox{\hbox
to0pt{\kern0.3\wd0\vrule height0.9\ht0\hss}\box0}}}}

\def\bbbs{{\mathchoice
{\setbox0=\hbox{$\displaystyle     \rm S$}\hbox{\raise0.5\ht0\hbox
to0pt{\kern0.35\wd0\vrule height0.45\ht0\hss}\hbox
to0pt{\kern0.55\wd0\vrule height0.5\ht0\hss}\box0}}
{\setbox0=\hbox{$\textstyle        \rm S$}\hbox{\raise0.5\ht0\hbox
to0pt{\kern0.35\wd0\vrule height0.45\ht0\hss}\hbox
to0pt{\kern0.55\wd0\vrule height0.5\ht0\hss}\box0}}
{\setbox0=\hbox{$\scriptstyle      \rm S$}\hbox{\raise0.5\ht0\hbox
to0pt{\kern0.35\wd0\vrule height0.45\ht0\hss}\raise0.05\ht0\hbox
to0pt{\kern0.5\wd0\vrule height0.45\ht0\hss}\box0}}
{\setbox0=\hbox{$\scriptscriptstyle\rm S$}\hbox{\raise0.5\ht0\hbox
to0pt{\kern0.4\wd0\vrule height0.45\ht0\hss}\raise0.05\ht0\hbox
to0pt{\kern0.55\wd0\vrule height0.45\ht0\hss}\box0}}}}

\def\bbbz{{\mathchoice {\hbox{$\sans\textstyle Z\kern-0.4em Z$}}
{\hbox{$\sans\textstyle Z\kern-0.4em Z$}}
{\hbox{$\sans\scriptstyle Z\kern-0.3em Z$}}
{\hbox{$\sans\scriptscriptstyle Z\kern-0.2em Z$}}}}

\begin{flushright}
\end{flushright}

\begin{center}
{\LARGE Temperature Independent Renormalization of \\
Finite Temperature Field Theory}
\end{center}

\begin{center}
{\Large \baselineskip 20pt
Christoph Kopper$^{\, a\,}$\footnote{email~: 
kopper@cpht.polytechnique.fr} $\,$,
\\
Volkhard F. M{\"u}ller$^{\, b\,}$\footnote{email~: 
vfm@physik.uni-kl.de} $\,$
\\
$\qquad$ and $\qquad$
\\
Thomas Reisz$^{\, c,d\,}$\footnote{email~: 
reisz@thphys.uni-heidelberg.de, supported by a Heisenberg Fellowship
}
}
\end{center}

\centerline{$^a$ Centre de Physique Th{\'e}orique de l'Ecole Polytechnique}
\centerline{F-91128 Palaiseau, France}
\vspace{0.1cm}
\centerline{$^b$ Fachbereich Physik, Universit{\"a}t Kaiserslautern}
\centerline{D-67653 Kaiserslautern, Germany}
\vspace{0.1cm}
\centerline{$^c$ Service de Physique Th{\'e}orique de Saclay, CE-Saclay}
\centerline{F-91191 Gif-sur Yvette Cedex, France}
\vspace{0.1cm}
\centerline{$^d$ Institut f{\"u}r Theor.~Physik, Universit{\"a}t Heidelberg}
\centerline{D-69120 Heidelberg, Germany}
\vspace{0.1cm}

\date{ }


\begin{abstract}
\normalsize
We analyse 4-dimensional
massive $\vp^4$ theory at finite temperature $T$ 
in the imaginary-time formalism. 
We present a rigorous proof that
this quantum field theory is renormalizable, to all orders
of the loop expansion. Our main point is to show that the 
counterterms can be chosen temperature independent, 
so that the temperature flow
of the relevant parameters as a function of $T$
can be followed. Our result confirms the experience
from explicit calculations
to the leading orders.
The proof is based on flow equations, 
i.e. on the (perturbative) Wilson renormalization group.
In fact we will show that the difference between the theories 
at $T>0$ and at $T=0$ contains no relevant terms. 
Contrary to BPHZ type formalisms
our approach permits to lay hand on renormalization 
conditions and counterterms
at the same time, since both appear as boundary terms of the
renormalization group flow. This is crucial for the proof.

\end{abstract}
\newpage

\section{Introduction }
Field theories at finite temperature and density have been proposed
as the fundamental underlying theory for the description
of the physics of the early universe.
A proposed scenario for baryogenesis is by the electroweak
phase transition \cite{linde}.
QCD is expected to become deconfined at high temperature.
The formation of a quark gluon plasma and the phase transitions
of QCD are supposed to be visible in relativistic
heavy ion collision and astrophysics
\cite{ortmanns}. A modern presentation of finite temperature
field theory can be found in \cite{lebellac}.

Beyond their phenomenological implications,
quantum field theories at finite temperature are
very challenging also from the more theoretical
point of view. 
There is a real-time as well as an imaginary-time formalism,
the first describing dynamical and the second equilibrium properties
\cite{landsman_vanweert}. 
Many fundamental issues and problems are unsolved so far
or require a deeper understanding.
Quantum field theories are subject to enhanced complexities
compared to zero temperature and zero density.
This is largely related to the presence of additional length scales,
due to the interaction with a heat bath.
On the various scales the properties of the theory are considerably
different.

The separation of scales is widely believed to be
an intrinsic property of the field theory.
In QCD the scales are associated to the generation of
electric and magnetic screening and plasmon masses.
In the framework of perturbation theory,
this manifests itself in terms of 
IR divergences that are ``severe''. 
They are not removable as it is the case at temperature $T=0$
by adjusting the renormalization prescription 
\cite{lowenstein}.
Various elaborate resummation techniques have been proposed to 
(at least partially) remove the IR singularities
and in addition compute screening masses
in perturbation theory.
In any case, all the approaches (need to) aim at a clean separation of
IR and UV behaviour.

A precondition of all these considerations is renormalizability.
Renormalizability is an essential requirement of any 
local quantum field theory, both at zero and non-zero temperature
\cite{zinn-justin}.
It implies that the correlation functions stay 
finite as the UV-cutoff $\Lambda_0$, say, is removed, 
$\Lambda_0\to\infty$,
and that the limit is parametrized by a set of renormalized 
(relevant) coupling constants.
Moreover, it is crucial that renormalization can be
achieved in a temperature independent way, 
which means that the field theory renormalized at zero temperature 
stays UV finite at every $T>0$ as well.
This is often taken for granted even for complicated theories,
such as gauge theories. Temperature independent renormalizability
is indispensable for relating 
bare and renormalized coupling constants in a 
$T$-independent way. 
It is thus required when formulating 
Callan-Symanzik type of equations that govern
the $T$-dependence of observables, including correlation functions
and effective masses. More generally it
implies that the static and dynamic properties
mediated by the interactions with a heat bath 
are intrinsic features of the field theory itself.

Various attempts and steps towards proving renormalizability
exist \cite{renorm}. 
In order to separate off the IR problem from the UV scale,
a massive field theory is considered.
Both in the real- and in the imaginary-time formalism,
the investigations are commonly based on a Feynman diagrammatic 
approach in momentum space.
In the real-time description,
it is argued that the part of the propagator which depends on
the temperature $T$ or the chemical potential $\mu$
decays exponentially fast for large momenta,
so it should be ``innocent'' of any UV problem.
In the imaginary-time formalism the approach is generically 
more ``cumbersome'', but it is again argued that
in the sum over the Matsubara frequencies
all $T$- or $\mu$-dependent UV divergences cancel out.

Experience obtained by explicit computations to
leading orders of perturbation theory confirms that,
once IR and UV singularities are properly disentangled,
all UV divergences 
found are $T$-independent and are removed by the zero temperature
 counterterms.
However, this is not so for non-zero chemical potential $\mu$
(associated to a finite density).
A field theory that has been renormalized at $\mu=0$ is able to generate
$\mu$-dependent UV divergences that are not removed by the
$\mu=0$ counterterms.
A simple example is given by a 4-dimensional
Yukawa model, with a chemical potential associated to the
fermion number.
In the framework of the renormalization group,
the chemical potential introduces an additional relevant operator,
so at least one additional renormalization
condition is expected.
This also indicates a possible problem for the analytic continuation from
the euclidean to the real-time formulation,
in agreement with a discussion  
\cite{steinmann} in the framework of axiomatic 
quantum field theories at finite temperature,
where the problem of proving the existence of correlation
functions (even at $\mu=0$) in the real-time formalism has been
stressed.

The renormalization of field theories at $T=0$ is well
understood. Strong statements and proofs 
on the renormalizability of various
field theories relevant in physics exist, including several different 
regularization and renormalization schemes, 
see e.g.~\cite{rigorous,kopper}.
Unfortunately, this sophistication does not extend to finite $T$ so far.
Rigorous proofs do not exist, to the best of our knowledge.
We would like to point out, however, that recently rigorous bounds,
uniform in the temperature, have been established for the perturbative
correlation functions of many-fermion models. Here renormalization is
necessary to obtain  well-behaved bounds on the IR side, 
when approaching the Fermi surface, whereas  
the UV regularization is kept fixed. Feldman et al.~\cite{feldman}
renormalize the many-fermion models with  $T\,$-independent
counterterms, as we do.   

In this paper we give a mathematical
proof that massive $\vp^4$ theory at finite $T$,
in the imaginary-time formalism, is renormalizable.
More precisely, we show,
to all orders of the loop expansion,
that all correlation functions become UV finite at every finite $T$
once the theory has been renormalized at $T=0$
by (one of the) usual renormalization prescriptions.

The proof is given in the framework of Wilson's flow equation.
It avoids the analysis of  
individual Feynman integrals (or Feynman sums),
which requires the involved combinatorics encoded in the forest formula
for overlapping divergences. Moreover it avoids the
formulation and proof of a 
power counting theorem.
Using flow equations, the proof of renormalizability merely
amounts to establish appropriate bounds in momentum space 
on the correlation functions, which are viewed
as coefficient functions of the associated
generating functional. The proof is
by induction on the number of loops.

This paper is organized as follows.
In Sect.~2 we introduce our basic notations.
This includes the definition of the generating functional 
$L^{\Lambda,\Lambda_0}(\vp)$ 
of the connected, free propagator amputated Green functions
on ``momentum scale $\Lambda$'', with $0\leq\Lambda\leq\Lambda_0$,
where $\Lambda_0$ denotes the UV cutoff.
The dependence of $L^{\Lambda,\Lambda_0}$ on the scale $\La$ 
is described by the so-called Wilson flow equation.
We recap the basic steps of proving renormalizability
of 4-dimensional $\vp^4$ field theories at zero temperature
by means of the flow equation.
Renormalizability is stated in terms of uniform bounds on the 
(coefficient functions of the) solution
$L^{\Lambda,\Lambda_0}(\vp)$ of the flow equation
and its derivative with respect to the UV-cutoff $\Lambda_0$,
with boundary conditions imposed
at $\Lambda=0$ for the relevant couplings and at $\Lambda=\Lambda_0$
for the irrelevant interactions.

In Sect.~3 we show that the difference 
$D^{\Lambda,\Lambda_0}(\uvp;T)$
of the generating functionals at temperature $T>0$ and $T=0\,$:
\eq
   D^{\Lambda,\Lambda_0}(\uvp;T) \equiv
     L^{\Lambda,\Lambda_0}(\uvp;T)-L^{\Lambda,\Lambda_0}(\uvp)
\label{one}
\eqe
has the properties of an irrelevant operator
in the sense of the renormalization group
\footnote{For the definition of the momentum space
field variables $\,\uvp\,$ and their position space Fourier
transform $\uhvp\,$ we refer to the beginning of sect.3~: 
Equ. (\ref{one}) should be understood in the weak sense, 
i.e. in a formal power series expansion w.r.t. $\hbar\,$
and as an identity for all coefficient functions generated
by the generating functionals. 
For the equation to make sense as it stands the variables
$\uhvp\,$ have to be appropriately restricted, for instance
to be smooth functions, supported in the interval 
$[0,\,\be]\,$ in the
$x_0$-component in position space.}. 
More precisely, $T$-independence of the counterterms 
means that the boundary condition
\eq \label{zero_T_counterterms}
   D^{\Lambda_0,\Lambda_0}(\uvp;T) \equiv 0
\eqe
holds. From this we derive strong bounds on all scales $\Lambda$ 
for $D^{\Lambda,\Lambda_0}(\uvp;T)\,$. 
Together with the bounds on $L^{\Lambda,\Lambda_0}(\vp)$
this proves UV finiteness of massive $\vp^4_4$ 
for every finite $T$, that is,
\eq
   \lim_{\Lambda_0\to\infty,\Lambda\to 0} L^{\Lambda,\Lambda_0}(\uvp; T)
\eqe
exists, to all orders of the loop expansion.
As an immediate consequence, the theory is also made UV finite 
by imposing normalization conditions on the mass, the wave function
constant and on the quartic coupling constant at any fixed temperature $T_0\,$.
In Sect.~4 we summarize our central statements
and give a short outlook.

\section{Renormalization of zero temperature $\vp_4^4$ theory\\ -
a short reminder}
Perturbative renormalizability of euclidean zero temperature $\vp_4^4$
theory 
will be established by analysing the generating functional
$L^{\La,\Lao}$ of connected (free propagator) amputated 
Green functions (CAG).
The upper indices  
$\La$ and $\Lao$ enter through the regularized propagator
\eq
C^{\La,\Lao}(p)\,=\, {1 \over  p^2+m^2} 
\{ e^{- {p^2+m^2 \over \Lao^2}} -e^{- {p^2+m^2 \over \La^2}} \}
\label{prop}
\eqe
or its  Fourier transform 
\eq
\hat{C}^{\La,\Lao}(x)\,=\, 
\int_p C^{\La,\Lao}(p) \,e^{ipx}\,,
\label{propx}
\eqe
where we use the shorthand 
\eq
\int_p~:= \int_{\bbbr^4}  {d^4 p \over (2\pi)^4}  \ .
\label{short1}
\eqe
We assume 
\eq
0 \le \La \le \Lao \le \infty
\eqe
so that the Wilson flow parameter $\La$ takes the role of an 
infrared (IR) cutoff\footnote{Such a cutoff is of course not necessary
in a massive theory. The IR behaviour is only modified for $\La$ above
$m$.}, whereas  $\Lao$ is the ultraviolet (UV)
regularization. The full propagator is recovered for 
$\La=0$ and $\Lao \to \infty\,$.
We also introduce the convention 
\eq
\hat{\vp}(x) = \int_p  \vp(p) \,e^{ipx}\,,\quad
{\de \over \de\hat{\vp}(x)} =
(2\pi)^4 
\int_p {\de \over \de \vp(p)}\, e^{-ipx}\ .
\label{short2}
\eqe
For our purposes the "fields" $\hat{\vp}(x)$ may be assumed to live in
the Schwartz space ${\cal S}(\bbbr^4)$.
For finite $\Lao$ and in finite volume the theory can be given
rigorous meaning starting from the functional integral
\eq
e^{- {1\over \hbar} (L^{\La,\Lao}(\hat\vp)+ I^{\La,\Lao})}
\,=\, 
\int \, d\mu_{\La,\Lao}(\hat\phi) \; 
e^{- {1\over\hbar} \Llol(\hat\phi\,+\,\hat\vp)} \ ,
\label{funcin}
\eqe
where the factors of $\hbar\,$ have been introduced to allow for a consistent
loop expansion in the sequel.
In (\ref{funcin}) $\,d\mu_{\La,\Lao}(\hat\phi) $ denotes the (translation
invariant) Gaussian measure with covariance $\hbar \hat{C}^{\La,\Lao}(x)$.
The normalization factor $\,e^{-{1\over\hbar} I^{\La,\Lao}} $ is
due to vacuum contributions. It
diverges in infinite volume so that we can take the infinite volume
limit only when it has been eliminated [10]. 
We do not make the finite
volume explicit here since it plays no role in the 
sequel.\footnote{A rigorous treatment of the thermodynamic
limit requires to replace the propagator
(\ref{propx}) by a finite volume version, e.g.
$\,
\hat{C}_V^{\La,\Lao}(x,y)\,=\,
\chi_V(x)\, \hat{C}^{\La,\Lao}(x-y)\,\chi_V(y)\;,
\,$
where $\chi_V\,$ is the characteristic function of the volume
$\,V$, and to regard the Gaussian measure with 
covariance
$\,\hat{C}_V^{\La,\Lao}(x,y)\,$. In this case the quantity
$\,I_V^{\La,\Lao}\,$ is obviously well defined, at any order $l\,$ in
$\,\hbar\,$. Then (\ref{feq}) is well-defined. 
After decomposing $\,L_V^{\La,\Lao}\,$ w.r.t. powers of $\hbar\,$ and 
of the field $\hat\vp\,$, we realize that the coefficient functions
$\,\cLlln \,$ are well-defined  in the thermodynamic limit, since
they are given as finite sums over UV-regularized connected diagrams. 
The existence of the thermodynamic limit is of course confirmed by the
bounds on the solutions of the FE. It should also be
feasible to study the thermodynamic limit itself with the aid of the FE
in finite volume, by proving inductively uniform bounds on the 
(appropriately defined) "translational
invariant part" of the finite volume Green functions and a convergence
statement analogous to  (\ref{propo2}).}

The functional $\Llol(\hat\vp)$ is the bare action including
counterterms, viewed as a formal power
series in $\,\hbar\,$. Its general form for 
symmetric\footnote{The necessary generalizations in the nonsymmetric
  case will be  surveyed in the end of the next section.} 
 $\vp_4^4$
theory is 
\[
   \Llol(\hat\vp) = {g \over 4!}  \int \! \! d^4 x \, \hat{\vp}^4(x)  
   \; + 
\]  
\eq
   + \int \! \!d^4 x \,\{{1 \over 2} a(\Lao)\hat{\vp}^2(x) +
    {1 \over 2} b(\Lao) \sum_{\mu=0}^3 (\pa_{\mu}\hat{\vp})^2(x) +
    {1 \over 4!}c(\Lao) \hat{\vp}^4(x)\} \, ,
\label{nawi}
\eqe  
where $g >0$ is the renormalized coupling, and   
the parameters $a(\Lao),\ b(\Lao),\ c(\Lao)$
fulfill
\eq
a(\Lao),\ b(\Lao),\ c(\Lao) =O(\hbar)\,.
\eqe
They are directly related to the standard mass, wave function and
coupling constant counterterms. 
Since in the flow equation framework it is not necessary to introduce
bare fields in distinction to renormalized ones 
(our field is the renormalized one in this language), there is a
slight difference, which is to be kept in mind only 
when comparing to other schemes. The Wilson flow equation (FE)  
is obtained from (\ref{funcin}) on differentiating 
w.r.t. $\La\,$. It is a differential equation for the functional
$L^{\La,\Lao}\,$~: 
\eq
\partial_{\La}(\Lll + I^{\La,\Lao} ) \,=\,\frac{\hbar}{2}\,
\langle\frac{\delta}{\delta \hat\vp},(\partial_{\La}\hat \Cll)
\frac{\delta}{\delta \hat\vp}\rangle\Lll
\,-\,
\frac{1}{2}\, \langle \frac{\delta}{\delta
  \hat\vp}\Lll,(\partial_{\La}
\hat \Cll)
\frac{\delta}{\delta \hat\vp} \Lll\rangle \;\,.
\label{feq}
\eqe
By $\langle\ ,\  \rangle$ we denote the standard scalar product in 
$L_2(\bbbr^4, d^4 x)\,$. Changing to momentum space and
expanding in a formal powers series w.r.t. $\hbar$ we write
with slight abuse of no\-tation
\eq
L^{\La,\Lao}(\vp)\,=\,\sum_{l=0}^{\infty} \hbar^l\,L^{\La,\Lao}_{l}(\vp)\,.
\label{3.3}
\eqe
{}From $L^{\La,\Lao}_{l}(\vp)$ we then obtain the CAG of loop order $l$
in momentum space  as 
\footnote{The  normalization of the ${\cLlln}$ 
is defined differently from earlier references.}
\eq
(2 \pi)^{4(n-1)} \de_{\vp(p_1)} \ldots \de_{\vp(p_n)}
L^{\La,\Lao}_l|_{\vp \equiv 0}
\,=\,
\de^{(4)} (p_1+\ldots+p_{n})\, {\cLlln}(p_1,\ldots,p_{n-1})\ ,
\label{cag}
\eqe
where we have written 
$\delta_{\vp(p)}=\delta/\delta\vp(p)$.
Note that our definition of the $\cLlln$ is such that $\cLll_{0,2}$
vanishes. The absence of 0-loop two (and one-) point functions is
important for the set-up of the inductive scheme, from which we will 
prove renormalizability below.
The FE (\ref{feq}) rewritten in terms of the CAG (\ref{cag})
takes the following form
\eq
\pa_{\La} \pa^w \,\cLlln (p_1,\ldots p_{n-1}) =
{1 \over 2} \int_k (\pa_{\La}\Cll(k))\,\pa ^w 
\cLll_{l-1,n+2}(k,-k, p_1,\ldots p_{n-1})
\label{fequ}
\eqe
\[
-\!\!\!\!\!\!\!\!\!\!\!\sum_{l_1+l_2=l,\,w_1+w_2+w_3=w\atop
n_1+n_2=n} \!{1 \over 2} 
\Biggl[ \pa^{w_1} \cLll_{l_1,n_1+1}(p_1,\ldots,p_{n_1})\,
(\pa^{w_3}\pa_{\La}\Cll(p'))\,\,
\pa^{w_2} \cLll_{l_2,n_2+1}(p_{n_1+1},\ldots,p_{n})\Biggr]_{ssym}\!\!\!,
\]
\[ \mbox{where }\quad
p'= -p_1 -\ldots -p_{n_1}\,= \,p_{n_1+1} +\ldots +p_{n}\ .
\] 
Here we have written (\ref{fequ})  directly in a form
where also momentum derivatives of the  CAG (\ref{cag})
are performed, and we used the shorthand notation
\eq
\pa^w:= \prod_{i=1}^{n-1}\prod_{\mu=0}^{3}
({\pa \over \pa p_{i,\mu}})^{w_{i,\mu}}\ \mbox{ with }\
w=(w_{1,0},\ldots,w_{n-1,3}),\  |w|=\sum w_{i,\mu}\,,\  w_{i,\mu}\in \bbbn_0\ .
\eqe 
The symbol $ssym$ 
\footnote{It is defined differently from the
symbol $sym$ in \cite{kopper}, the present conventions being slightly
more elegant.} means summation over those permutations of the momenta 
$p_1,\ldots, p_n$, which do not leave invariant the subsets
$\{p_1,\ldots, p_{n_1}\}$ and $\{p_{n_1+1},\ldots,p_{n}\}$.
Note that the CAG are symmetric  in their momentum arguments by definition.
A simple inductive proof of the renormalizability of $\vp_4^4$ theory 
has been exposed several times in the literature \cite{kopper}, 
and we will not repeat it in detail. 
The line of reasoning can be resumed as follows.\\
The induction hypotheses to be proven are~:\\
A) Boundedness
\eq
|\pa^w \cLlln(\vec{p})| \le\,
(\La+m)^{4-n-|w| }\,{{\cal P}_1}(log {\La + m \over m})\,
{{\cal P}_2}({|\vec{p}| \over \La+m})\ .
\label{propo1}
\eqe
B) Convergence
\eq
|\pa_{\Lao} \pa^w \cLlln(\vec{p})| \le\,
{1\over \Lao^2} (\La+m)^{5-n-|w| }\,{{\cal P}_3}(log {\Lao  \over m})\,
{{\cal P}_4}({|\vec{p}| \over \La+m})\ .
\label{propo2}
\footnote{ In fact, in symmetric $\vp_4^4$ theory ${1\over \Lao^2} $
can be replaced by ${\La\over \Lao^3} $ as shown in \cite{KoPe}.}
\eqe
Here and in the following the ${\cal P}$ denote (each time they appear
possibly new) polynomials with nonnegative coefficients. 
The coefficients 
depend on $l,n,|w|,m$,
but not on $\vec{p},\,\La,\,\Lao$. We used the shorthand
$\vec{p}=(p_1,\ldots,p_{n-1})$ and $|\vec{p}|=\sup
\{|p_1|,\ldots,|p_n|\}$.
The statement (\ref{propo2}) implies renormalizability, 
since it proves that the limits
$\,\lim_{\Lao \to \infty,\;\La \to 0}  \cLlln(\vec{p})\,$ exist
to all loop orders $l\,$.
But the statement (\ref{propo1}) has to be obtained first
to prove (\ref{propo2}). 
The inductive scheme to prove the statements proceeds upwards in $l$,
for given $l$ upwards in $n$, and for given $(l,n)$ downwards in $|w|$, 
starting from some arbitrary $|w_{max}| \ge 3$. 
The important point to note is that the terms on the r.h.s. of the FE
always are prior to the one on the l.h.s. in the inductive order.
So the bound (\ref{propo1}) may be used as an induction hypothesis on
the r.h.s. Then we may integrate the FE, where terms with $n+|w| \ge
5$ are integrated down from $\Lao$ to $\La$, since for those terms we
have the boundary conditions following from (\ref{nawi})
\eq
 \pa^w \,\cLlol_{l,n} (p_1,\ldots p_{n-1}) =0\,,\, \mbox{ for}\quad n+|w| >4\,,
\label{bo1}
\eqe
whereas the terms with $n+|w| \le 4$ at the renormalization point - 
which we choose at zero momentum for simplicity -      
are integrated upwards from $0$ to $\,\La$, since they are fixed by
($\Lao$-independent)
renormalization conditions, fixing the relevant parameters of the
theory\footnote{The simplest choice would be to set 
$a^R_l=0,\  b^R_l=0,\ c^R_l =0\,$, in which case the renormalized
coupling is identical to the connected four point function
at zero momentum. A shift away from zero  momentum 
would result in nonvanishing terms $c^R_l\, $, just to mention
one example of more general choices.}~:
\eq
\cLol_{l,2} (p) = a^R_l + b^R_l \,p^2 +O((p^2)^2) \,, \quad 
\cLol_{0,4} (0) = g\,,\ 
\cLol_{l,4} (0) = c^R_l\,,\ l\ge 1 \, .
\label{bo2}
\eqe
Symmetry considerations tell us that there are no other
nonvanishing renormalization constants apart from $a ^R_l,\ b^R_l,\  c^R_l\,$, 
and the Schl{\"o}milch or integrated Taylor formula permits us
to move away from the renormalization point, treating first 
$\cLol_{l,4}$ and then the momentum derivatives of
$\cLol_{l,2}\,$, in descending order. With these remarks on the
boundary conditions, and using the bounds on the propagator  
and its derivatives
\eq
|\pa^w \pa_{ \La}
C^{\La,\Lao}(p)| \leq\,
\La^{-3-|w| }\,{\cal P}(|p|/\La) \; e ^{-{p^2+m^2 \over \La ^2}}\;,
\label{probo}
\eqe
statement A) above is straightforwardly verified by inductive
integration of the FE. Once this has been achieved statement B)
follows on applying the same inductive scheme to bound the solutions 
of the FE, integrated over $\La$ and then derived w.r.t. $\Lao\,$. 

\section{ Temperature independent renormalization of\\ 
finite temperature $\vp_4^4$ theory}
We fix the notations recalling at the same time some basic facts about
euclidean finite temperature field theory. The scalar field $\hat\vp(x)$  
becomes periodic in $\,x_0\,$ at finite temperature with period
$\be=1/T$. Correspondingly position space integrals 
over the zero component of the coordinates are now restricted 
to the compact interval $[0,\,\be]\,$. 
Symbols denoting finite temperature quantities will
generally be underlined, thus we write
\eq
\up~:=(\up_0,\,\vecp)~:= (2\pi n T,\,\vecp)\,,\ n \in \bbbz\,,\
\,\, \int_{\up}:= T \,\sum_{n \in \bbbz}\, \int_{\bbbr^3} 
{d^3p \over (2\pi)^3}\ .
\label{short3}
\eqe
We also introduce the convention 
\eq
{\uhvp}(x)~:= \int_{\up}  \uvp(\up)\, \,e^{i\up x}\,,
\quad
\uvp(\up) \,=\int_0^{\be} dx_0 \int_{\bbbr^3} d^3x \;\uhvp(x)
\,\,e^{-i\up x}\,,
\label{short4}
\eqe
\eq
{\de \over \de{\uhvp}(x)} =
{(2\pi)^3 \over T} 
\int_{\up} {\de \over \de \uvp(\up)}\, e^{-i \up x}\,,\quad
{\de \over \de{\uvp}(\up)} =
{T \over (2\pi)^3 } 
\int_0^{\be} dx_0 \int_{\bbbr^3} d^3x \,
 {\de \over \de \uhvp(x)}\, e^{i \up x}\ .
\label{short5}
\eqe
The regularized propagator now takes the form 
\eq
C^{\La,\Lao}(\up)\,=\, {1 \over  \up^2+m^2} 
\{ e^{-{ \up^2+m^2 \over \Lao^2}} -e^{- {\up^2+m^2 \over \La^2}} \}\ .
\label{propt}
\eqe
The generating functional 
of the finite temperature CAG will
be called $\Lll(\uvp;T)$. In analogy with (\ref{cag}) we define the
CAG  through  
\eq
\de_{\uvp(\up_1)} \ldots \de_{\uvp(\up_n)}
L^{\La,\Lao}_l(\uvp;T)|_{\uvp \equiv 0}
\,=\,
\eqe
\[
({T\over (2\pi)^3})^{n-1}\,\,
\large \de_{0,(\up_{1,0}+\ldots+\up_{n,0})}\
\de^{(3)} (\vecp_1+\ldots+\vecp_{n})\, {\cLlln}(\up_1,\ldots,\up_{n-1};T)\ .
\label{cagt}
\]
At this stage we could prove renormalizability of the finite
temperature theory in the same way as for the zero temperature
theory. A slight difference is 
that the relations (\ref{bo2}) are to be replaced by
\[
\cLol_{l,2} (\up; T) = a^R_l(T) + b^{R,0}_l(T) \; \up_0^2 +
b^{R,1}_l(T) \; \vec p^{\,2} +\,O(\up^4) \, ,
\]
\eq
\cLol_{0,4} (\up=0; T) = g\,,\ \cLol_{l,4} (\up=0; T) = c_l^R(T) 
\,,\ l\ge 1 \,, 
\label{bo3}
\eqe
since the space-time $O(4)$-symmetry is broken down to a $\bbbz_2 \times
O(3)$-symmetry which demands a new renormalization condition. 
However we want to go beyond and prove
temperature independent renormalizability, in the sense
that the counterterms can be chosen temperature independent. To do so,
it is advantageous to pass directly to the difference between      
the finite and zero temperature theories, which we will do now. 
Note in this respect that if we wanted to prove the
renormalizability of the finite temperature theory, keeping the
counterterms fixed at their zero temperature values, would not work
within our scheme and procedure~: The CAG would become arbitrarily
divergent in $\Lao$ with increasing loop order, 
since integrating relevant terms from
$\Lao$ to  0 (instead of integrating them from a renormalization
condition fixed at $\La=0\,$ up to $\Lao\,$) gives divergent integrals.   
Thus we rather study the difference functions 
\eq
\Dlln(\{\up\})~:= \cLlln(\{\up\};T)\,-\,\cLlln(\{\up\}) \ .
\label{diff}
\eqe
We only define and need the difference CAG 
$\,\,\Dlln$ at the external
momenta $(\{\up\}):=(\up_1,\ldots,\up_{n-1})$.
{}From the FE (\ref{fequ}) and the analogous equation for the
$\cLlln(\{\up\};T)\,$ we can derive a FE for the $\Dlln(\{\up\})$ in
the following form~:
\eq
\pa_{\La}  \,\Dlln (\{\up\}) \,=\,\!\!\!
{1 \over 2} \int_{\uk} (\pa_{\La}\Cll(\uk))\,
\Dll_{l-1,n+2}(\uk,-\uk,\{\up\})
\label{fequd}
\eqe
\[
+\,{1 \over 2} \biggl\{
\int_{\uk} (\pa_{\La}\Cll(\uk))\, 
\cLll_{l-1,n+2}(\uk,-\uk,\{\up\})\,-\,
\int_{k} (\pa_{\La}\Cll(k))\, 
\cLll_{l-1,n+2}(k,-k,\{\up\}) \biggr\}
\]
\[
-\!\!\!\sum_{l_1+l_2=l,\atop
n_1+n_2=n} {1 \over 2} 
\Biggl\{
\Biggl[  \cLll_{l_1,n_1+1}(\up_1,\ldots,\up_{n_1};T)
(\pa_{\La}\Cll(\up'))\,\,
\Dll_{l_2,n_2+1}(\up_{n_1+1},\ldots,\up_{n})\Biggr]_{ssym}\!\!
\]
\[
+\Biggl[ \Dll_{l_1,n_1+1}(\up_1,\ldots,\up_{n_1})
(\pa_{\La}\Cll(\up'))\,\,
\cLll_{l_2,n_2+1}(\up_{n_1+1},\ldots,\up_{n})\Biggr]_{ssym}
\Biggr\}\ ,
\]
where again
\[
\up'= -\up_1 -\ldots -\up_{n_1}\,=\, \up_{n_1+1} +\ldots +\up_{n}\;.
\] 
The boundary conditions we want to impose on the system $\Dlln$
are (from the previous remarks) obviously the following ones~: 
\eq
\Dlol_{l,n} (\up_1,\ldots, \up_{n-1}) \,= \,0 \,, \quad l,n \in \bbbn\;.
\label{bo6}
\eqe
To start the induction we also note 
\eq
\Dll_{0,n}(\up_1,\ldots, \up_{n-1})\,=\,0\,, \quad n \in \bbbn \,,  
\label{tree}
\eqe
at the tree level all difference terms $\Dll_{0,n}$ vanish.
This follows from the fact that restricted to the momenta
$(\up_1,\ldots \up_{n-1})$ the tree level functions 
$\cLll_{0,n}(\up_1,\ldots \up_{n-1};T)$ and 
$\cLll_{0,n}(\up_1,\ldots \up_{n-1})$ agree. 
Now we would like to use the same inductive scheme 
proceeding upwards in $l$, and for given $l$ upwards in $n$, 
to prove the finiteness of $\lim_{\Lao \to \infty,\La \to 0} \Dll_{0,n}\,$.
Due to the form of (\ref{bo6}) 
we {\it always} integrate the FE for $\Dlln$ from
$\Lao$ down to $\La$, since the boundary terms at $\Lao\, $
always vanish. We want to prove the following\\[.1cm] 
{\it Theorem}~: 
\eq
|\Dll_{l,n}(\up_1,\ldots, \up_{n-1})|\,\le\,
(\La+m)^{-s-n}\,{{\cal P}_1}(\log {\La + m \over m})\,
{{\cal P}_2}({|\{\up\}| \over \La+m})\,,
\label{propo3}
\eqe
\eq
|\pa_{\Lao} \Dll_{l,n}(\up_1,\ldots, \up_{n-1})|\,\le\,
{1 \over \Lao ^2} (\La+m)^{-s-n}\,{{\cal P}_3}(log {\Lao \over m})\,
{{\cal P}_4}({|\{\up\}| \over \La+m})\ .
\label{propo4}
\eqe
The nonnegative coefficients in the  polynomials  ${\cal P}$ 
depend on $l,n,s,m$ and (smoothly) on $T$, 
but not on $\{\up\},\,\La,\,\Lao\,$. The positive integer
$s \in \bbbn$ may be chosen arbitrarily.\\
The finite temperature CAG $\cLll_{0,n}(\up_1,\ldots, \up_{n-1};T)\,$,
when renormalized with the {\it same} counterterms as the zero temperature
ones, satisfy the same bounds as in (\ref{propo1},\ref{propo2})
restricted to $w=0\,$. The
coefficients in the polynomials ${\cal P}$  
may now depend on $l,n,m$ and (smoothly) on $T\,$. 
\\[.1cm] 
{\it Remark}~: It is possible to prove the bounds
(\ref{propo1},\ref{propo2}) also for derivatives of
the finite temperature CAG $\cLll_{0,n}(\up_1,\ldots, \up_{n-1};T)\,$.
In the $p_{i,0}$-components differentiations then have to replaced by finite
differences. However these bounds are not needed in the inductive proof,
so we skip them here.\\[.1cm]
{\it Proof}~: 
We first prove (\ref{propo3}) and and the statement corresponding to 
(\ref{propo1}) for $w=0\,$, using the inductive 
scheme indicated previously. 
Regarding the FE (\ref{fequd}) we state that it is
compatible with the inductive scheme and that the only term in which
 (\ref{propo3}) cannot be used as an induction hypothesis is the 
following one~:
\eq 
\int_{\uk} (\pa_{\La}\Cll(\uk))\, 
\cLll_{l-1,n+2}(\uk,-\uk,\{\up\})\,-\,
\int_{k} (\pa_{\La}\Cll(k))\, 
\cLll_{l-1,n+2}(k,-k,\{\up\}) \ .
\label{pdif}
\eqe
So our sharp $\La$-bound on  $\Dlln\,$ 
can only be verified if it holds for this difference term.
Here we use (\ref{propo1},\ref{propo2}) and the
Euler-MacLaurin-formula, see e.g. \cite{Bou}.
We can rewrite (\ref{pdif}) as 
\eq
{-2 \over \La^3} \int {d^3\vec k \over (2\pi)^4}\, 
e^{-{ \vec k^2+m^2 \over \La^2}}\, \biggl[ 2\pi T \sum_{n \in \bbbz} 
g(2\pi n T)-
\int_{-\infty}^{\infty} dk_0\, g(k_0)\biggr]\;,  
\label{ema1}
\eqe
where we introduced the function
\eq
 g(k_0)\,=\,e^{- {k_0^2 \over \La^2}}\;
\cLll_{l-1,n+2}(k,-k,\{\up\})\,
\ \mbox{ for } \,\,\vec k,\, \{\up\}\, \mbox{ fixed}\ .
\label{ema2}
\eqe
The Euler-MacLaurin formula for our case can be stated in the form
\eq
2\pi T \sum_{n \in \bbbz} g(2\pi n T)-
\int_{-\infty}^{\infty}  dk_0\, g(k_0)  \,=\,
-\pi T \,[ g(\infty)-g(-\infty)]
\label{ema3}
\eqe
\[
\,+\,
\sum_{k=1}^{r+1}  {b_{2k} (2\pi T)^{2k} \over (2k)!} 
[g^{(2k-1)}(\infty)-g^{(2k-1)}(-\infty)]\,+\, R_{r+1}\ .
\]
Here $b_{2k}\,$ are the Bernoulli numbers. We observe that 
passing to the limit of an infinite integration interval
is justified, since the function $g(k_0)$ and its derivatives vanish
rapidly at infinity. The remainder $R_{r+1}$ obeys the following bound
 \cite{Bou}
\eq
| R_{r+1}| \le \, 4\, e^{2\pi} T^{2r+3} \int_{-\infty}^{\infty} dk_0 
\,|g^{(2r+3)}(k_0)|\;,
\label{ema4}
\eqe
therefore we obtain, using again (\ref{propo1},\ref{propo2}) 
\eq
| R_{r+1}| \le \, T^{2r+3} \,{(\La+m)^{2-n} \over \La^{2r+2} }\,
{{\cal P}_1}(log {\La + m \over m})\,
{{\cal P}_2}({|\{k,\up\}| \over \La+m})\ .
\label{ema5}
\eqe
Note that $r \in \bbbn$ can be chosen arbitrarily here, and the bound
for (\ref{pdif}) is thus 
\eq
T^{2r+3} \,e^{-{m^2 \over \La ^2} } \,{(\La+m)^{2-n} \over \La^{2r+2} }\,
{{\cal P}_1}(log {\La + m \over m})\,
{{\cal P}_2}({|\{k,\up\}| \over \La+m})
\label{ema6}
\eqe
\[
\le \,T^{2r+3} \,(\La+m)^{2-n-2r-2} \,
{{\cal P}_3}(log {\La + m \over m})\,
{{\cal P}_4}({|\{k,\up\}| \over \La+m})\ .
\]

After this preparation we consider the induction process~:
At each loop order we first prove (\ref{propo3}), and then 
(\ref{propo1}) for finite $T$ and corresponding momenta. 
This second step is trivial from (\ref{propo1},\ref{propo2})  
at $T=0$, from the definition (\ref{diff}) and from (\ref{propo3})
\footnote{We may choose the bounds for $s=0$ from 
(\ref{propo3},\ref{propo4}) when  bounding the finite temperature CAG,
so that polynomials appearing in the bounds may be chosen s-independent.}.
We know already the theorem to
be true at 0 loop order. This and the form of  the FE (\ref{fequd}) 
implies that we do not  need a bound on any of the $\cLll_{l,n}(\{\up\};T)$ 
in the inductive bound on $\Dlln$ at the given loop order $l$.

It is instructive to regard how the induction starts at loop order
$l=1$. Treating first the case $n=2$ we find that the only
nonvanishing 
contribution on the r.h.s. of the FE stems from (\ref{pdif}), and it
is momentum independent, so that
integrating over $\La$ we get 
\[
|\Dll_{1,2}(\up)| \le \, c\,(\La+m)^{-2r-1} 
\]
with a suitable constant $c\,$, depending on $r\,$.
For $n=4$ also the last two terms on the r.h.s. of the FE contribute. 
Using the result for $\Dll_{1,2}(\up)\,$, integration over $\La$ gives
\[
|\Dll_{1,4}(\{\up\})| \le \,(\La+m)^{-2-2r-1} \,
{{\cal P}}({|\{\up\}| \over \La+m})\ .
\]
{}From this one inductively obtains the bound for $n\ge 6$
\[
|\Dll_{1,n}(\{\up\})| \le \,(\La+m)^{-(n-2)-2r-1} \, 
{{\cal P}}({|\{\up\}| \over \La+m})\ .
\]
Having bounded the difference functions $\Dll_{1,n}$ we can bound the 
CAG $\,\cLll_{1,n}(T) =\cLll_{1,n}(T\!\!=\!\!0)+ \Dll_{1,n}\,$, 
see (\ref{diff}). 
Then we may proceed inductively to higher loop orders 
and verify the inductive bound
\[
|\Dll_{l,n}(\{\up\})| \le \,(\La+m)^{-(n-2)-2r-1} \, 
{{\cal P}_1}(\log {\La + m \over m})\,
{{\cal P}_2}({|\{\up\}| \over \La+m})\ .
\]
This proves the first part of the
theorem on writing $s=2r-1\,$ for $s$ odd, and majorizing to
obtain even $s$. It follows that
the $\cLll_{l,n}(T)\,$ may be bounded in agreement
with (\ref{propo1},\ref{propo2}).

Now we turn to the proof of the statement (\ref{propo4}) which implies
convergence of the $\Dll_{l,n}$ for $\Lao \to \infty\,$.
The proof is based on the same inductive scheme and starts from the FE
(\ref{fequd}) integrated over $\La\,$ from $\Lao\,$ to $\La\,$, and then
derived w.r.t. $\Lao\,$. The result is of the form 
\eq
-\pa_{\Lao} \,\Dlln (\{\up\}) \,=\,
\mbox{[RHS of  (\ref{fequd})]}|_{\La=\Lao}\,+\, 
\int_{\La}^{\Lao} d\la\; \pa_{\Lao}  \mbox{[RHS of  (\ref{fequd})]}(\la)\,,  
\label{lao}
\eqe
and we denote the RHS of this equation shortly as 
\[
I_{l,n}^{\Lao}(\{\up\})\,+\,I_{l,n}^{\La,\Lao}(\{\up\})\,\,.
\]
Since we have imposed $\,\cLlol_{l,n}(T) \equiv \cLlol_{l,n}\,$,
and since moreover these terms vanish for $n \ge 6$, we find 
\eq
I_{l,n}^{\Lao}(\{\up\})\,= -\de_{n,2}\,\,
\Biggl[
\int_{\uk} { e ^{- {\uk^2+m^2 \over \Lao ^2}} \over \Lao ^3}
-\int_{k} { e ^{- {k^2+m^2 \over \Lao ^2}} \over \Lao ^3}
\Biggr]\, \cLlol_{l-1,4}\ .
\label{i}
\eqe
Since $\;\cLlol_{l-1,4}\equiv c_{l-1}(\Lao)\,,\,\ l>1\,$ and
$\;\cLlol_{0,4}\equiv g\,\,$,
see (\ref{nawi}), we realize that (\ref{i}) is momen\-tum independent.
The difference can be calculated explicitly or bounded again  
using the Euler-MacLaurin formula, and we obtain
\eq
|I_{l,n}^{\Lao}|\le\,
\de_{n,2}\,
\Lao ^{-2-2r}\,\,
{{\cal P}}(\log {\Lao \over m})
\eqe
for $r \in \bbbn$ and a suitable ${{\cal P}}$ depending on $r$. 
To get a bound on $I_{l,n}^{\La,\Lao}(\{\up\})\,$ 
we apply the derivative in (\ref{lao}) to all entries using
the product rule (noting that when applied to 
$\pa_{\La} C^{\La,\Lao}$ it gives zero). In any case the derivative
brings down the required  factor of $\Lao ^{-2}$, either by
(\ref{propo2}), or by (\ref{propo4}) together with
the induction hypothesis. Apart from this the bound 
(\ref{propo4}) is obtained similarly as (\ref{propo3}), 
using in particular the Euler-MacLaurin formula for the difference 
term (\ref{pdif}) derived w.r.t. $\Lao\,$. 
This proves also (\ref{propo4}).
\qed

We end this section with two remarks on possible generalizations. 
First the preceding analysis can be extended to 
nonsymmetric $\vp_4^4$-theory. The action
(\ref{nawi}) then has to be replaced by
\eq
\ti L^{\Lao,\Lao}(\hat\vp) = 
\Llol(\hat\vp)\,+\,{h \over 3!} \int\! d^4 x \, \hat{\vp}^3(x)  
\,+\,\int \!d^4 x \,\{ {1 \over 3!} \,d(\Lao)\,\hat{\vp}^3(x)+
v(\Lao) \, \hat{\vp}(x)\}
\label{nawii}
\eqe  
with the tree level three-point coupling $h$ and $\Lao$-dependent
parameters 
\eq
d(\Lao)\,,\ v(\Lao) \,=\,O(\hbar)
\eqe
implementing the counterterms necessary to renormalize the one- and
three-point functions.
Correspondingly we pose additional renormalization conditions
\eq
\cLol_{l,1}  = v^R_l \,, \quad 
\cLol_{l,3} (0) = d^R_l \,\,\,\, \mbox{ for } \,\, l\in \bbbn\ \;, 
\label{bzr}
\eqe
to be joined to (\ref{bo2}). Then the bounds (\ref{propo1},\ref{propo2})
hold again, but are no more trivially fulfilled for $n\,$
odd.\footnote{These bounds can be improved by replacing $n$ by 
$\hat n$, defined to be the smallest even integer greater or equal to $n\,$.}
Once the theory at $T=0\,$ is bounded, the differences (\ref{diff})
again yield the theory at $T>0\,$. Bounds corresponding to  
(\ref{propo3},\ref{propo4}) are proven proceeding as before, in the
symmetric case.

As a second remark, we point out
that for the existence of the large cutoff
limit $\Lambda_0\to\infty$,
it is not necessary that the relevant coupling constants
are subject to normalization conditions at zero temperature.
Equally well we can impose normalization conditions
at some temperature $T_0>0\,$. We pointed out that 
at finite temperature the space-time $O(4)$-symmetry is broken
down to $\bbbz_2 \times O(3)\,$. Then 
the 3 independent renormalization
constants $a^R$, $b^R$ and $c^R$ at $T=0$, (\ref{bo2}), 
become replaced by four parameters
$a^R(T_0)$, $b^{R,0}(T_0)$, $b^{R,1}(T_0)$ and $c^R(T_0)$ at $T_0$,
cf.~(\ref{bo3}),
corresponding to four relevant couplings.
However, starting from an $O(4)$-symmetric 
zero temperature theory we have proved that 
\eq
   L^{\Lambda,\Lambda_0}(\uvp;T_0)-L^{\Lambda,\Lambda_0}(\uvp)
\eqe
has the properties of an irrelevant operator.
This implies that for given $b^{R,0}(T_0)\,$ there is a unique choice 
for  $b^{R,1}(T_0)\,$, or vice versa, such that the finite temperature 
theory stems from an $O(4)$-symmetric zero temperature theory. 
Any different choice would be associated to a 
zero temperature theory, where $O(4)$-symmetry is broken by hand 
through the renormalization conditions. 
Note that the $O(4)$-symmetric choice is generally not the one 
where $b^{R,0}(T_0)=b^{R,1}(T_0)\,$: Integration over $\La\,$, 
starting from the same counterterms (the $O(4)$-symmetric ones)
will lead to a finite difference at $\La=0\,$, since $O(4)$-invariance 
is broken in the propagator. Otherwise stated, the fact that the 
finite temperature theory stems from an $O(4)$-symmetric zero
temperature theory, can be simply recognized on inspection of the
counterterms, but not on the renormalization conditions.

\section{ Summary }

We have presented a proof for the perturbative renormalizability of
massive finite temperature $\vp_4^4\,$-theory. The starting point are the 
bounds (\ref{propo1},\ref{propo2}) which prove the renorma\-lizability 
of the zero temperature theory. In the flow equation framework they
serve at the same time as induction hypotheses for the inductive
proof. Bounds of this type have by now been rigorously established for
nearly all theories of physical interest, including gauge theories,
where the restoration of the Ward identities in the final theory pose 
an additional problem, to be solved by a suitable restriction on the
renormalization conditions. Taking due care of the exceptional momentum  
problem, corresponding bounds can also be established for theories with
massless particles.
  
To extend the  bounds to  the corresponding finite temperature theories  
presents no really new problems for the practitioner.  
The main problem to be solved rather is that 
the existence of the correlation functions in the large cutoff limit
should be proven without changing the counterterms.
In our setup this corresponds to posing
the boundary conditions (\ref{bo6})  
for the difference Green functions $\cal D\,$ between the $T\!>0\,$ 
and the $T=0\,$ theories. The anounced result is contained in the bounds
(\ref{propo3},\ref{propo4}). The main new technical tool 
used to get there is the Euler-MacLaurin formula, 
generalized to an infinite integration interval for a rapidly
decaying integrand.
It is applied to
the difference terms appearing in the flow equations for the
functions $\cal D\,$ 
that are not bounded by the induction hypothesis alone,
(see (\ref{pdif})- (\ref{ema6})). Here it
comes to our help that the bounds  (\ref{propo1},\ref{propo2}) are
sufficiently powerful so as to transform momentum derivatives into
negative powers of $\La\,$. Via the Euler-MacLaurin formula it is then
possible to gain an arbitrary power in $\La\,$ paying the
corresponding power in $T\,$ (see \ref{ema5}). This
achieves (far more than) showing that all difference functions
$\cal D$ are irrelevant. For the latter a gain of a power of $\La
^{2+\vep}$ would have sufficed. We emphasize again that our result
agrees with the experience and intuition gained from 
explicit perturbative calculations.

Renormalization is a central issue that is strongly related
to the fundamental principles of local quantum field theory.
Renormalizability of a field theory gives it a
meaning beyond some low energy effective model.
The techniques we have presented here for proving
renormalizability of a field theory at finite temperature
mainly rely on two properties.
The first property is renormalizability at zero temperature.
The second one is that
the difference between the theory at finite and zero temperature
act like an irrelevant operator
that does not spoil renormalizability.
Renormalization group flow equations provide an appropriate
tool to put this statement on a strong basis and
prove renormalizability for finite temperature.
We expect that these methods generalize appropriately
to apply to more realistic and complex field theories such as QCD,
where both the UV and the IR scale problem are to be attacked.


\end{document}